\documentclass[journal]{IEEEtran}

\usepackage{cite}

\usepackage[normalem]{ulem}

\usepackage[pdftex]{graphicx}
  \graphicspath{{images/}}

\usepackage{color}

\usepackage{amsmath}
\usepackage{mathtools}
\usepackage{amsfonts}
\usepackage{amssymb}
\usepackage{amsthm}
\usepackage{bbm}

\usepackage[font=footnotesize,skip=3pt]{caption}

\usepackage{algorithm}
\usepackage{algorithmic}
\usepackage{array}
\usepackage{url}
\usepackage{lipsum}
\usepackage{acro}

\usepackage{tikz}

\usetikzlibrary{arrows, arrows.meta, shapes, shapes.multipart, trees, positioning, backgrounds, fit, matrix, petri, calc, automata}

\usepackage{ellipsis}

\usetikzlibrary{calc}

\usetikzlibrary{decorations.pathreplacing,decorations.markings,shapes.geometric}

\makeatletter
\def\BState{\State\hskip-\ALG@thistlm}
\makeatother

\tikzset{radiation/.style={{decorate,decoration={expanding waves,angle=90,segment length=4pt}}}}

\tikzset{cross/.style={cross out, draw=black, minimum size=2*(#1-\pgflinewidth), inner sep=0pt, outer sep=0pt},
	cross/.default={4pt}}

\usepackage[absolute,showboxes]{textpos}

\TPMargin{5pt}

\begin{document}
\title{Meteorologically Introduced Impacts on Aerial Channels and UAV Communications 
\vspace{0.2cm}}
\author{authors
\thanks{M. Song, Y. Huo, T. Lu, and Z. Liang are with the Department of Electrical and Computer Engineering, University of Victoria, BC, Canada. Z. Liang is with Chang'an Univeristy, Xi'an, China. 
}

\vspace{-0.4cm}} 

\author{\IEEEauthorblockN{Mengan Song\textsuperscript{1,2},
Yiming Huo\textsuperscript{2}, Tao Lu\textsuperscript{2}, Xiaodai Dong\textsuperscript{2}, Zhonghua Liang\textsuperscript{1}}\\

\thanks{M. Song and Z. Liang's work was supported in part by the Fundamental Research Funds for the Central Universities, CHD under Grant 300102249303, and in part by the Natural Science Basic Research Program in Shaanxi Province of China under Grant 2020JM-242.}

\IEEEauthorblockA{
\textsuperscript{1}School of Information Engineering, Chang'an University, Xi'an 710064, China \\
Email: 2018024005@chd.edu.cn, lzhxjd@hotmail.com \\
\textsuperscript{2}Department of Electrical and Computer Engineering, University of Victoria, Victoria BC V8W 3P6, Canada \\
Email: ymhuo@uvic.ca and \{taolu, xdong\}@ece.uvic.ca}
}

\maketitle

\begin{abstract}
As 5G wireless systems and networks are now being globally commercialized and deployed, more diversified application scenarios are emerging, quickly reshaping our societies and paving the road to the beyond 5G (6G) era when terahertz (THz) and unmanned aerial vehicle (UAV) communications may play critical roles. In this paper, aerial channel models under multiple meteorological conditions such as rain, fog and snow, have been investigated at frequencies of interest (from 2 GHz to 900 GHz) for UAV communications. Furthermore, the link budget and the received signal-to-noise ratio (SNR) performance under the existing air-to-ground (A2G) channel models are studied with antenna(s) system considered. The relationship between the 3D coverage radius and UAV altitude under the influence of multiple weather (MW) conditions is simulated. Numerical results show that \textcolor{black}{medium rain has the most effects on the UAV's coverage} for UAV communications at millimeter wave (mmWave) bands, while snow has the largest impacts at near THz bands. In addition, when the frequency increases, the corresponding increase in the number of antennas can effectively compensate for the propagation loss introduced by weather factors, while its form factor and weight can be kept to maintain the UAV's payload.

\end{abstract}


\IEEEpeerreviewmaketitle
\vspace{-0.3cm}
\section{Introduction}
In recent years, with  rapid advancements of artificial intelligence (AI) technologies and the mass production of low-cost drones, unmanned aerial vehicle (UAV) communications have become a promising and necessary integral part of the development of 5G and beyond networks \cite{Zhang:UAV-Cellular}, \cite{Huo:UAV-IoT}, \cite{Huo:UAV}, and Internet-of-Things (IoT) applications \cite{Huo:Maritime}. In general, UAV-enabled wireless communications can be categorized to three main use cases, namely, UAV-aided ubiquitous coverage, relaying, and information dissemination and data collection \cite{Zeng:Accessing}. Moreover, the former two categories are suitable for 5G and beyond aerial base station (BS) data offloading \cite{Huo:VTC2019} and wireless connectivity relaying. 

For example, UAVs can be used to cope with  emergency communications during executing the rescue missions in the middle of (and/or after) the natural disasters, such as, earthquakes, mudslides, floods, forest fires, etc. Furthermore, in the case of power outage, UAVs can achieve fast communication and service recovery thanks to its flexible and quick deployment. UAV can function as the aerial base station (ABS) for future networks, to address massive data connection and transmission, such as real-time ultra-high-definition (UHD) video streaming and relaying \cite{Zeng:Accessing}.
\par However, one of the practical application challenges confronting 5G and beyond UAV communications is to adopt multiple weather scenarios, which has not been widely studied in previous works. To be more specific, despite some limited studies on meteorological influence on radar and wireless communications \cite{Zang:The Impact}-\cite{Zhao:Rain attenuation}, there is very little research on how weather conditions could affect UAV communications for 5G and beyond networks. More specifically, although the sub-6 GHz carrier frequencies mainly used for 4G and 5G low/mid bands do not produce significant absorption, reflection and scattering, in terms of their radio propagation characteristics \cite{Rappaport:5G}, employing 5G high bands ($>$ 20 GHz) and even THz bands (possibly used for 6G and beyond \cite{Rappaport:6G}) can be largely affected by the various meteorological conditions (e.g., the atmospheric precipitation) due to the decrease of the wavelength on top of the gaseous attenuation.

In this paper, we first study the characteristics of radio propagation at a wide range of representative frequencies from 2 GHz up to 900 GHz under various meteorological conditions, namely, rain, fog, and snow. Secondly, we conduct a detailed review of the air-to-ground (A2G) channel models \cite{Hourani:Optimal LAP}-\cite{Hourani:Modeling air-to-ground}, which unveils the relationship between the optimal height of UAV and the maximum coverage. The meteorology constrained A2G channel models are further derived in standard atmosphere, on top of International Telecommunication Union (ITU) models of rain and fog  \cite{ITU:rain}-\cite{ITU:fog} and Oguchi's model of snow \cite{Oguchi:Electromagnetic}, at the frequencies of interest in this research. Furthermore, in order to address the future 5G/6G UAVs communication challenges from the aforementioned weather-constrained aerial channels (including both A2G and air-to-air (A2A) channel models), this paper investigates the received signal-to-noise ratio (SNR) under various application scenarios with multiple antennas (antenna arrays) that enable the compensation of the path loss increment, respectively. Finally, we propose practical system-level design considerations. To our best knowledge, this is the first work that has investigated meteorological impacts on aerial communications on a very wide range of frequencies.


\vspace{-0.25cm}
\section{characteristics of radio propagation under various meteorological conditions}
\label{sec:system_model}

One of the most drastic challenging issues lies in the weather conditions for deploying UAVs for telecommunications services, since the real-life climate and weather are geographically diverse, and can change dramatically even within one specific location. In order to lay the foundation of utilizing UAV communications under such dynamic environments and conditions, specific models of multiple typical weather conditions first need to be carefully investigated.  

\subsection{Specific Attenuation Model for Rain}
\vspace{-0.0cm}
\label{sec:a}

In terms of ITU model, rainfall-introduced specific attenuation is $\gamma_\text{R}$ (dB/km) is related to the power law of the rainfall intensity $R$ (mm/h), and is written as \cite{ITU:rain}
\begin{equation}
\begin{aligned}
\label{eqn:gammarain}
\gamma_\text{R}=k R^{\alpha} ,
\end{aligned}
\end{equation}
where $k$ and $\alpha$ are determined by the following equations that are derived by fitting the curve obtained from the experimental discrete data to the power-rate coefficient,

\begin{equation}
\begin{aligned}
\label{eqn:logk}
\log _{10} k=\sum_{j=1}^{4}\left(a_{j} \exp \left[-\left(\frac{\log _{10} f-b_{j}}{c_{j}}\right)^{2}\right]\right)
\\+m_{k} \log _{10} f+c_{k} , 
\end{aligned}
\end{equation}

\begin{equation}
\begin{aligned}
\label{eqn:alpha}
\alpha=\sum_{j=1}^{5}\left(a_{j} \exp \left[-\left(\frac{\log _{10} f-b_{j}}{c_{j}}\right)^{2}\right]\right)\\
+m_{\alpha} \log _{10} f+c_{\alpha} ,
\end{aligned}
\end{equation}
where $f$ is the frequency in GHz, and $a_{j}$, $b_{j}$, $c_{j}$, $m_{k}$, $c_{k}$, $m_{\alpha}$, and $c_{\alpha}$ are fitting values. $k_{\text{H}}$ and $\alpha_{\text{H}}$ which will be used later stand for the specific antenna horizontal polarization coefficients while $k_{\text{V}}$ and $a_{\text{V}}$ represent constant values of vertical polarization coefficients. 

For all path geometries in linear and circular polarization, the coefficients in (\ref{eqn:gammarain}) can be calculated from (\ref{eqn:k}) and (\ref{eqn:a}) using the values given in (\ref{eqn:logk}) and (\ref{eqn:alpha}) as

\begin{equation}
\begin{aligned}
\label{eqn:k}
k=\left[k_{\text{H}}+k_{\text{V}}+\left(k_{\text{H}}-k_{\text{V}}\right) \cos ^{2} \theta_{\text{p}} \cos 2 \tau\right] / 2 ,
\end{aligned}
\end{equation}
\begin{equation}
\begin{aligned}
\label{eqn:a}
a=\left[k_{\text{H}} a_{\text{H}}+k_{\text{V}} a_{\text{V}}+\left(k_{\text{H}} a_{\text{H}}-k_{\text{V}} a_{\text{V}}\right) \cos ^{2} \theta_{\text{p}} \cos 2 \tau\right] / 2 k,
\end{aligned}
\end{equation}
where $\theta_{\text{p}}$ is the path oblique angle, 
and $\tau$ is the polar oblique angle of the relative horizontal position. 

\subsection{Specific Attenuation Model for Fog}
\vspace{-0.0cm}
\label{sec:a}
According to the ITU model, fog is another frequent meteorological phenomenon that can largely affect the UAV communications and flight. The specific amount of attenuation in the fog can be expressed as \cite{ITU:fog}
\begin{equation}
\begin{aligned}
\label{eqn:gammafog}
&\gamma_{\text{c}}(f, T)=K_{\text{l}}(f, T) M\quad(\mathrm{dB}/\mathrm{km}), 
\end{aligned}
\end{equation}
where $\gamma_c$ represents the specific attenuation in the fog (dB/km), $T$ is the temperature of liquid water, $K_\text{l}$ is the specific attenuation coefficient 
((dB/km)/($\mathrm{g}/\mathrm{m}^{3})$) give in (\ref{eqn:fogkl}) and $M$ is the density of liquid water in the cloud or fog ($\mathrm{g}/\mathrm{m}^{3}$). Fog attenuation can be very significant at frequencies around 100 GHz or above. For medium fog, the density of liquid water in the fog is usually about 0.05 $\mathrm{g}/ \mathrm{m}^{3}$ (visibility is about 300 m), and dense fog is 0.5 $\mathrm{g}/ \mathrm{m}^{3}$ (visibility is about 50 m)\cite{ITU:fog}.  
\begin{equation}
\begin{aligned}
\label{eqn:fogkl}
K_{\text{l}}(f, T)=\frac{0.819 f}{\varepsilon^{\prime \prime}\left(1+\eta^{2}\right)} \quad(\mathrm{dB} /\mathrm{km})/\left(\mathrm{g}/ \mathrm{m}^{3}\right),
\end{aligned}
\end{equation}
where $  \eta=\frac{2+\varepsilon^{\prime}}{\varepsilon^{\prime \prime}} $ and the complex permittivity of water can be expressed as
\begin{equation}
\begin{aligned}
\label{eqn:varepsilon2}
\varepsilon^{\prime \prime}(f)=\frac{f\left(\varepsilon_{0}-\varepsilon_{1}\right)}{f_{p}\left[1+\left(f / f_{p}\right)^{2}\right]}+\frac{f\left(\varepsilon_{1}-\varepsilon_{2}\right)}{f_{s}\left[1+\left(f / f_{s}\right)^{2}\right]} ,
\end{aligned}
\end{equation}
\begin{equation}
\begin{aligned}
\label{eqn:varepsilon1}
\varepsilon^{\prime}(f)=\frac{\varepsilon_{0}-\varepsilon_{1}}{\left[1+\left(f / f_{p}\right)^{2}\right]}+\frac{\varepsilon_{1}-\varepsilon_{2}}{\left[1+\left(f / f_{s}\right)^{2}\right]}+\varepsilon_{2} ,
\end{aligned}
\end{equation}
where $\varepsilon_{0}=77.66+103.3(\theta-1)$, $\varepsilon_{1}= 0.0671 \varepsilon_{0} $,  $\varepsilon_{2}= 3.52$, $\theta=300/ T_{\text{fog}}$, $T_{\text{fog}}$ is the temperature in fog weather set to $293.15K$, while $f_{p}$ and $f_{s}$ are the primary and secondary relaxation frequencies $f_{p}=20.20-146(\theta-1)+316(\theta-1)^{2}$, $f_{s}=39.8f_{p}$ $(\mathrm{GHz})$.

\subsection{Specific Attenuation Model for Snow}
\vspace{-0.0cm}
\label{sec:a}
The specific attenuation of propagation caused by snow is based on the model in\cite{Oguchi:Electromagnetic}.  In terms of the specific liquid content, snow can be divided into dry snow and wet snow \cite{Koch:snow}, and we only consider the model of dry snow given by
\begin{equation}
\begin{aligned}
\label{eqn:gammasnow}
\gamma_{\text{s}}=0.00349 \frac{R_{\text{s}}^{1.6}}{\lambda^{4}}+0.00224 \frac{R_\text{s}}{\lambda} \quad(\mathrm{dB}/\mathrm{km}) ,
\end{aligned}
\end{equation}
where $R_{\text{s}}$ is the snowfall speed in millimeters per hour and $\lambda$  is the wavelength in centimeters. 

\subsection{Attenuation by Atmospheric Gases}
\vspace{-0.0cm}

Considering the air contains oxygen, nitrogen, rare gases and water vapor that absorb radio waves and cause atmospheric attenuation, this effect needs to be formulated and introduced into the free space path loss model to obtain the propagation attenuation of electromagnetic waves in the standard atmosphere. In  \cite{ITU:gas}, the atmospheric attenuation $\beta$  is calculated as
\begin{equation}
\begin{aligned}
\label{eqn:beta}
\beta&=\beta_{o}+\beta_{w}
\\&=0.1820 f\left(N_{\text {Oxygen }}^{\prime \prime}(f)+N_{\text {Water Vapour}}^{\prime \prime}(f)\right)\quad(\mathrm{dB}/\mathrm{km}),
\end{aligned}
\end{equation}
where $\beta_{o}$ and  $\beta_{w}$ are the specific attenuation (dB/km) due to dry air (oxygen, pressure-induced nitrogen) and water vapour, respectively, and $N_{\text {Oxygen }}^{\prime \prime}(f)$ and $N_{\text {Water Vapour}}^{\prime \prime}(f)$ are the imaginary parts of the frequency-dependent complex refractivities respectively \cite{ITU:gas}.

\subsection{Propagation Model under Multiple Weather Conditions}
\vspace{-0.0cm}
Combining the specific propagation attenuation of rain, fog and snow in the previous subsections with the free-space propagation model at a standard atmosphere, the path loss under multiple weather (MW) conditions for wireless channels is obtained as
\begin{equation}
\begin{aligned}
\label{eqn:plmw}
PL_{\text{MW}}=32.442+20 \lg (f)+20 \lg (d)+\beta d+\gamma d ,
\end{aligned}
\end{equation}
where $f$ is the frequency in MHz, $d$ is the distance between the transmitter and receiver in kilometers, $\beta$ represents specific gaseous attenuation and $\gamma$ is the specific attenuation caused by rain, fog or snow, which are denoted by $\gamma_{\text{R}}$,     $\gamma_{\text{c}}$ and $\gamma_{\text{s}}$ in dB per kilometer (km). \textcolor{black}{Based on (\ref{eqn:plmw}), some numerical results can be obtained as follows.}
    \begin{figure}[t]
		\centering
		\includegraphics[width=0.95\columnwidth]{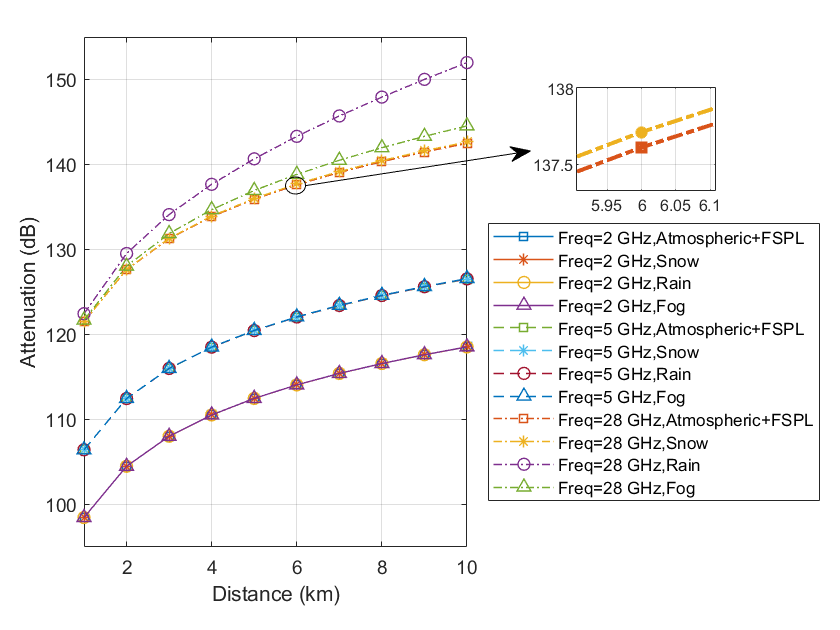}
		\caption{Ultra high frequency band propagation loss in multiple weather conditions.}
		\label{fig:pl_2_5_28}
		\vspace{-0.4cm}
	\end{figure}
	
	\begin{figure}[t!]
		\centering
		\includegraphics[width=0.95\columnwidth]{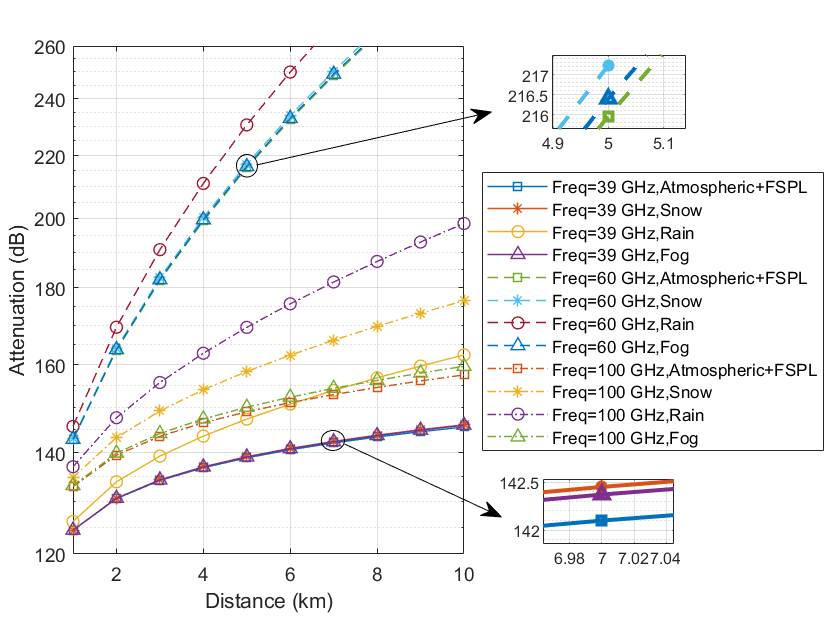}
		\caption{Extremely high-frequency band propagation loss in multiple weather conditions.}
		\label{fig:pl_39_60_100}
		\vspace{-0.4cm}
	\end{figure}
	
	\begin{figure}[t!]
		\centering
		\includegraphics[width=0.95\columnwidth]{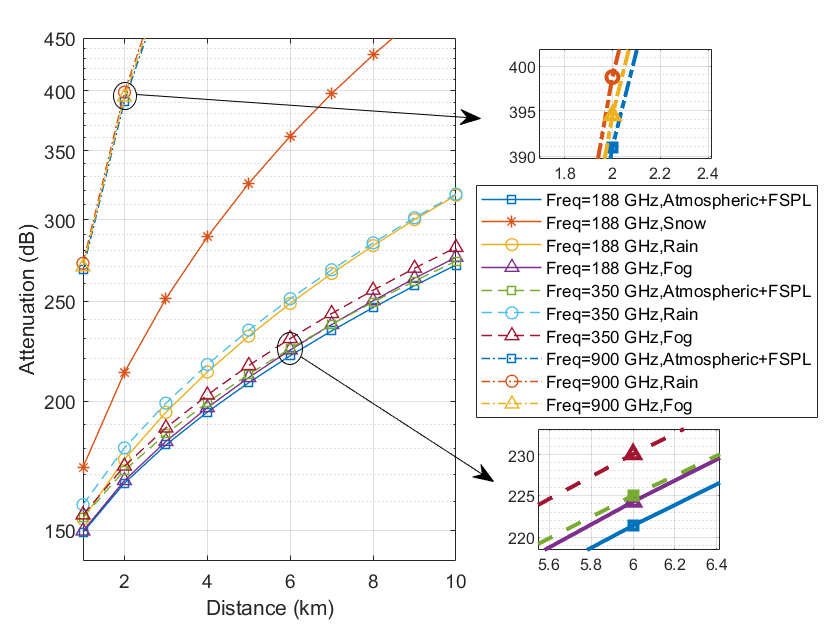}
		\caption{Low THz band propagation loss in multiple weather conditions.}
		\label{fig:pl_188_350_900}
		\vspace{-0.4cm}
	\end{figure}
As illustrated in Fig.~\ref{fig:pl_2_5_28}, the propagation attenuation differences of rain, fog and snow are not significant until ultra-high frequency (centimeter wave (cmWave)) bands. Fig.~\ref{fig:pl_39_60_100} shows the multiple weather propagation loss in the extremely high frequency (millimeter wave (mmWave)) bands. As compared to Fig.~\ref{fig:pl_2_5_28}, it can be seen from Fig.~\ref{fig:pl_39_60_100} that rain, fog and snow cause different degrees of propagation loss. Specifically, in the extreme high frequency band, the propagation attenuation under rain, snow and fog is listed with descending order. Moreover, snow and fog have similar propagation attenuation in relatively low frequency bands, for example at 39 GHz. In addition, compared to 39 GHz and 100 GHz, multiple weather induced attenuation at 60 GHz are particularly obvious. This is due to the serious absorption of electromagnetic waves at 60 GHz by oxygen in the air. Moreover, as illustrated in Fig.~\ref{fig:pl_188_350_900}, when the frequency is close to 1 THz, the propagation attenuation difference of rain, fog and snow changes significantly compared to Fig.~\ref{fig:pl_39_60_100}, in which snow has the greatest impact on propagation attenuation, followed by rain, and fog has the least impact on propagation attenuation. It's worth noting that the attenuation of snow is not shown at 350 and 900 GHz, due to enormous snow propagation attenuation and steep specific attenuation curve \cite{Oguchi:Electromagnetic}, while the specific attenuation curves for fog and rain are relatively lower.



\vspace{-0.2cm}
\section{Aerial channel models under various meteorological conditions}
\label{sec:problem}
In this section, we first describe the A2G channel model for UAV communications, and then the attenuation effects caused by rain, fog and snow are respectively derived for the A2G channel. The received SNR performance of practical UAV communication systems under multiple weather conditions is studied, with adaptive number of antennas proposed to compensate the path loss in THz bands.

\subsection{Conventional A2G Model without Atmosphere and Weather Impacts}
\vspace{-0.0cm}
\label{sec:a}
The A2G model used in this section can find the optimal height $h$ corresponding to maximizing the UAV coverage radius $R$. According to \cite{Hourani:Optimal LAP},  probabilities at line of sight (LoS) and at non-line-of-sight (NLOS) are constant given by
\begin{equation}
\begin{aligned}
\label{eqn:P_los_nlos}
P_{\text{LoS}}=&\frac{1}{1+a\exp (-b(\frac{180}{\pi }\tan^{-1}(\frac{h}{r})-a))} ,\\
P_{\text{NLoS}}=&1-P_{\text{LoS}} ,
\end{aligned}
\end{equation}
where $h$ is the altitude of UAV, $a$ and $b$ are environment dependent variables. The path loss for LoS and NLoS can be written as \cite{Hourani:Optimal LAP}
\begin{equation}
\begin{aligned}
\label{eqn:pl_los_nlos}
PL_{\text{LoS}}=20\log (\frac{4\pi f d}{c})+\eta _{\text{LoS}},\\PL_{\text{NLoS}}=20\log (\frac{4\pi f d}{c})+\eta _{\text{NLoS}} ,
\end{aligned}
\end{equation}
where $f$ is the carrier frequency, $c$ is the speed of light and $d$ denotes the distance between UAV and user given by $d=\sqrt{h^{2}+r^{2}}$. Moreover, $\eta _{\text{LoS}}$ and $\eta _{\text{NLoS}}$ are the environment dependent average additional path loss for LoS and NLoS conditions respectively. Based on (\ref{eqn:P_los_nlos}) and (\ref{eqn:pl_los_nlos}),  and consider the atmospheric attenuation $\beta$,  the standard or conventional A2G channel model can be obtained as
\begin{equation}
\begin{aligned}
\label{eqn:PL1}
PL_{\text{aerial}}=&PL_{\text{LoS}}\times P_{\text{LoS}}+PL_{\text{NLoS}}\times P_{\text{NLoS}}+\beta \times \frac{d}{1000} ,
\end{aligned}
\end{equation}

\subsection{Modified A2G Channel Models for Different Meteorological Conditions}
\vspace{-0.0cm}
\label{sec:a}
Combining the specific attenuation of rain, fog and snow with this model, the A2G channel models for UAV communications under multiple weather conditions can be derived as
\begin{equation}
\begin{aligned}
\label{eqn:PL}
PL_{\text{aerial}\text{MW}}=&(PL_{\text{LoS}}\times P_{\text{LoS}}+PL_{\text{NLoS}}\times P_{\text{NLoS}})+\frac{(\beta+\gamma)d}{1000} \\
=&(\frac{A}{1+a\exp (-b(\frac{180}{\pi }\tan^{-1}(\frac{h}{r})-a))}\\
&+20\log \frac{r}{\cos (\frac{180}{\pi }\tan^{-1}(\frac{h}{r}))}+B)+\frac{(\beta+\gamma)d}{1000},
\end{aligned}
\end{equation}
where $A=\eta _{\text{LoS}}-\eta _{\text{NLoS}}$, $B = 20\log (\frac{4\pi f}{c})+\eta _{\text{NLoS}}$, and $\gamma$ represents the specific attenuation coefficients of the multiple weather scenarios in $\mathrm{dB}/\mathrm{km}$.

\subsection{Aerial Communication Link Budget and Antenna(s) System Design}
\vspace{-0.0cm}
\textcolor{black}{For the link budget analysis, both transmitter and receiver ends are assumed to use antenna arrays (which depends on frequency bands, form factor, etc.).} Furthermore, considering the UAV payload is limited, the dimension of the entire antenna system is set to a fixed value, and we set the dimension of the patch antenna to 10 cm $\times$ 10 cm. The center-to-center spacing between the antenna elements is set to   $\frac{\lambda}{2}$ and the size of one-single antenna element is $\frac{{\lambda}_{\text{e}}}{2}$ where ${\lambda}_{\text{e}}$ denotes the effective wavelength that can be calculated as \cite{Huo:5G Equipment}
\begin{equation}\label{eq:WAVE}
{{\lambda }_{\text{e}}}\text{=}\frac{{{c}_{0}}}{f\sqrt{{{\varepsilon}_{\text{e}}}}}\ ,
\end{equation}
where $c_{\text{0}}$ is the speed of light in vacuum, $f$ is the frequency, and ${\varepsilon}_{\text{e}}$ is the effective dielectric constant that makes the effective wavelength shorter. The maximum number of antenna elements within the given area can be derived as  
\begin{equation}
\begin{aligned}
\label{eqn:antN}
N_{\text{ant}}=&(W\times\frac{2}{\lambda}+\frac{{\lambda}_{\text{e}}}{\lambda}-1)^{2},
\end{aligned}
\end{equation}
where $W$ is the length of the antenna array which is set to 10 cm. Based on (\ref{eqn:antN}), we can obtain values of antenna gain ($G_{\text{ant}}$) at transmitter and receiver, respectively, denoted by $G_{\text{ar}}$ and $G_{\text{at}}$. That is, $G_{\text{ant}}=G_{\text{ar}}=G_{\text{at}}$, and $G_{\text{ant}}=4+10\times\log_{10}(N_{\text{ant}})$. 

    \begin{figure}[t!]
		\centering
		\includegraphics[width=1.0\columnwidth]{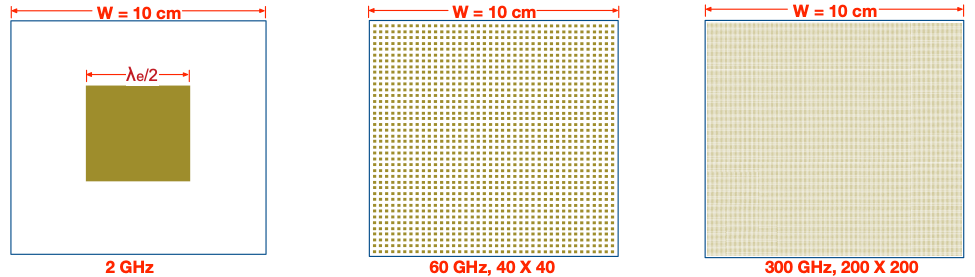}
		\caption{Comparison of antenna and antenna arrays within fixed area, for UAV communications at 2 GHz, 60 GHz and 300 GHz, respectively.}
		\label{fig:antenna}
		\vspace{-0.4cm}
	\end{figure}

As illustrated in Fig.~\ref{fig:antenna}, when the given physical dimension is fixed at 10 cm $\times$ 10 cm, a single patch antenna working at 2 GHz and patch antenna arrays operating at 60 GHz (40 $\times$ 40) and 300 GHz (200 $\times$ 200) are compared. In particular, the antenna arrays' gain of 60 GHz and 300 GHz are simulated as 36 dB and 50 dB, respectively. Then we can derive the signal power at receiver end, which is represented by $P_{\text{r}}$,

\begin{equation}
\begin{aligned}
\label{eqn:pr}
P_{\text{r}}=P_{\text{t}}-L_{\text{fr}}-L_{\text{ft}}+G_{\text{ar}}+G_{\text{at}}-F_{\text{s}} ,
\end{aligned}
\end{equation}
where the transmit power $P_\text{t}$ is set to 45 dBm  \cite{FCC}, $L_{\text{fr}}$ and  $L_{\text{ft}}$ represent the front-end loss of the receiver and the transmitter, which are both set to 1 dB. In addition, the noise can be calculated as \cite{Zang:The Impact}
\begin{equation}
\begin{aligned}
\label{eqn:noise}
N=&10\times\log _{10}(K\times T \times B)+ NF +30 ,
\end{aligned}
\end{equation}
where $K$ is Boltzmann constant, $T$ represents temperature ($T=298.15$ K), and $NF$ denotes the noise figure, which is set to 1 dB for 2 GHz, 2 dB for 60 GHz, and 6.5 dB for 300 GHz \cite{Tessmann:LNA}. According to (\ref{eqn:pr}) and (\ref{eqn:noise}), the received SNR can be obtained, $SNR=P_{\text{r}}-N$.


\section{numerical results}
\label{sec:algorithm}

\begin{figure}[t!]
		\centering
		\includegraphics[width=0.9\columnwidth]{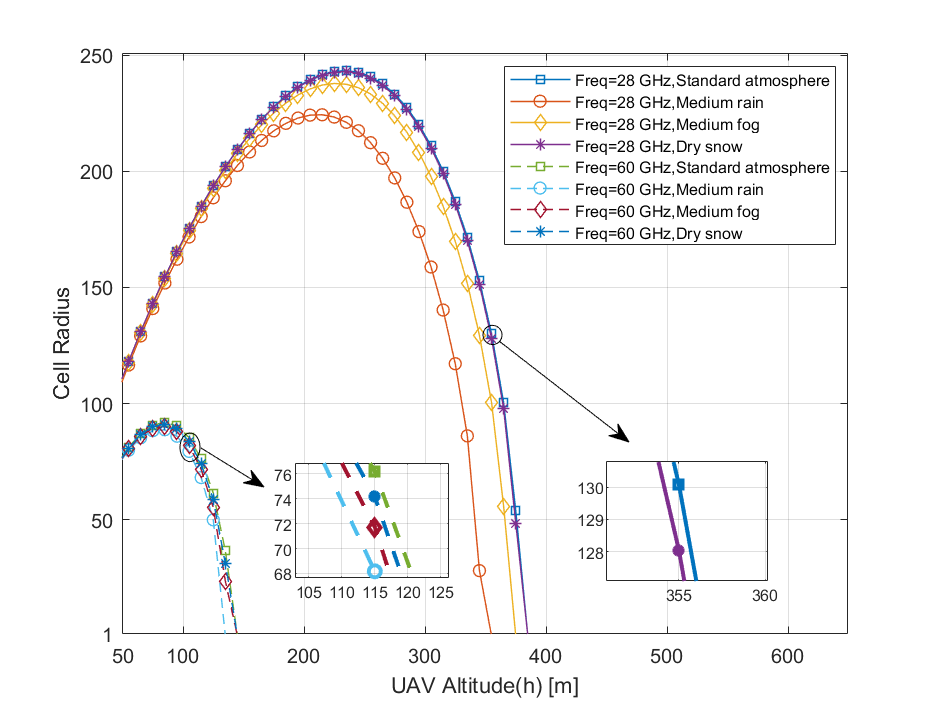}
		\caption{Cell radius versus UAV altitude for mmWave bands under rain, fog, and snow with various levels.}
		\label{fig:uav_28_60}
		\vspace{-0.4cm}
	\end{figure}

\begin{figure}[t!]
		\centering
		\includegraphics[width=0.9\columnwidth]{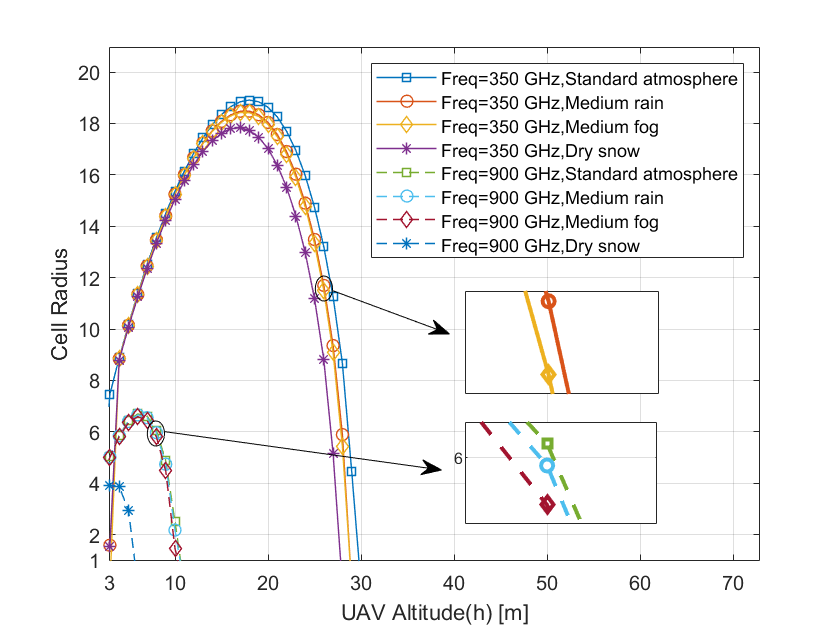}
		\caption{Cell radius versus UAV altitude for THz bands under rain, fog, and snow with various levels.}
		\label{fig:uav_350_900}
		\vspace{-0.4cm}
	\end{figure}

\begin{figure}[t!]
	\centering
		\includegraphics[width=0.9\columnwidth]{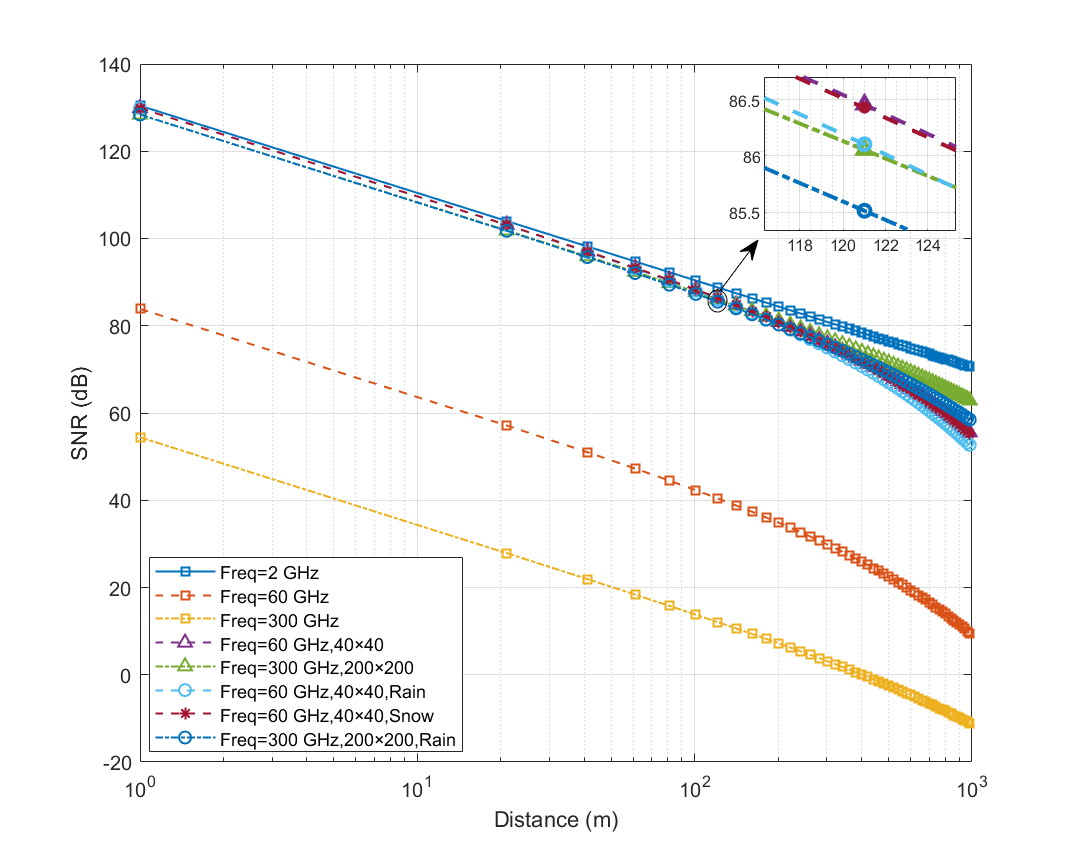}
		\caption{SNR versus distance at 2 GHz, 60 GHz and 300 GHz with different antenna(s) configurations.}
		\label{fig:snr}
		\vspace{-0.4cm}
	\end{figure}

 
In this section, we evaluate the A2G channel models based on multiple weather introduced impacts. The evaluation is mainly composed of two parts, the first one, based on Figs.~\ref{fig:uav_28_60} and \ref{fig:uav_350_900}, presents numerical results of A2G propagation loss of UAV communications in multiple weather conditions. We set $R=12.5$ mm/h (medium rain), $M=0.05$ $\mathrm{g}/\mathrm{m}^{3}$ (medium fog) and $R_{\text{s}}$ in Fig.~\ref{fig:uav_28_60} is set to 5 mm/h (snow), while $R_{\text{s}}$ in Fig.~\ref{fig:uav_350_900} is 0.5 mm/h (snow). The second part of numerical results includes Figs.~\ref{fig:antenna} and \ref{fig:snr}, in which the dimension of the antenna system is fixed to 10 cm $\times$ 10 cm, $P_\text{t} =45$ dBm \cite{FCC}. 

Figs.~\ref{fig:uav_28_60} and \ref{fig:uav_350_900} show multiple weather impacts on cell radius versus altitude of UAV communications in  mmWave bands and terahertz (THz) bands respectively with standard atmosphere. As illustrated in Fig.~\ref{fig:uav_28_60}, at 28 GHz and 60 GHz in the mmWave bands, rain, fog and snow have different degrees of influence on the coverage radius and height of UAV communication. Specifically, when it is moderate rain, the coverage radius and UAV height have the smallest values, while when it is light snow, we have the largest coverage radius and UAV height. The medium fog induced impact is in between. However, when it comes to  near THz bands, the UAV coverage radius during light snow is the smallest as illustrated in Fig.~\ref{fig:uav_350_900}. In other words, compared with rain and fog, snow has the most severe propagation loss in UAV communication at THz bands. In addition, impacts of medium rain and medium fog on UAV communication in the THz band are not much different.

As further illustrated in Fig.~\ref{fig:snr}, the received SNR versus distance is given for carrier frequencies at 2 GHz, 60 GHz, 300 GHz, respectively. When antenna arrays are configured at mmWave and THz frequencies, compensation can be observed to enhance link budget, under average rain and snow at mmWave band and average rain at low THz band. However, snow caused attenuation at THz is too huge as observed from Fig.~\ref{fig:pl_188_350_900} and therefore it is not discussed for compensation using regular-dimension antenna arrays here.      

\vspace{-0.25cm}
\section{conclusion}
\label{sec:system_model}
In this paper, we have investigated the aerial channel models constrained by multiple weather conditions based on the specific attenuation models from ITU and Oguchi. Based on the mathematical models, the propagation attenuation of rain is most severe for mmWave bands while the severity of attenuation caused by snow grows and surpasses rain for THz bands. In addition, results show that adopting adaptive number of antenna elements can effectively improve the SNR without introducing payload punishment. Adaptive beamforming schemes for UAV mmWave/THz communications with meteorological impacts can be studied as future work.



\end{document}